| | |
|---|---|
| Title | **Preparation and characterization of ultra-hydrophobic calcium carbonate nanoparticles** |
| Authors | A Barhoum[1,2], S M El-Sheikh[3], F Morsy[2], S El-Sherbiny[2], F Reniers[4], T Dufour[4], M P Delplancke[5], G Van Assche[1], H Rahier[1] |
| Affiliations | [1] Department of Materials and Chemistry, Vrije Universiteit Brussel, Pleinlaan 2, 1050 Brussels, Belgium<br>[2] Department of Chemistry, Helwan University, 11795 Helwan, Egypt<br>[3] Nanostructured Materials Division, Advanced Materials Department, CMRDI, P. O. Box 87, Helwan, 11421, Cairo, Egypt<br>[4] Analytical and Interfacial Chemistry, Université Libre de Bruxelles (ULB), Brussels, Belgium<br>[5] Department 4MAT, Universite Libre de Bruxelles, 50 avenue F.D. Roosevelt, 1050 Bruxelles, Belgium |
| Ref. | IOP Conf. Series: Materials Science and Engineering, 2014, Vol. 64, 012037 |
| DOI | http://dx.doi.org/10.1088/1757-899X/64/1/012037 |
| Abstract | Anionic surfactants based on fatty acids are usually used to modify the particle surface properties of $CaCO_3$ with the aim to enhance its dispersion and compatibility with polymer matrices. In this study sodium oleate was used for the preparation of ultra-hydrophobic $CaCO_3$ nanoparticles using a wet carbonation route. The effect of sodium oleate on the characteristics, particle size, morphology, surface potential, thermal decomposition and hydrophobicity of $CaCO_3$, was investigated using X-ray diffraction (XRD), Fourier transform infrared spectroscopy (FTIR), transmission electron microscopy (TEM), Zeta potential, thermogravimetric analysis (TGA) and water contact angle measurement (WCA). The results showed that the addition of 2 wt% sodium oleate helps in reducing the particle size from 2 μm length scalenohedral particles to 45 nm rhombohedral particles and modifying of the hydrophobic property of $CaCO_3$. |

# 1. Introduction

Calcium carbonate nanoparticles have received much attention owing to its wide industrial applications such as paper, rubber, plastics, and paints industries. $CaCO_3$ produces at industrial-scale via wet carbonation route by bubbling $CO_2$ gas through an aqueous $Ca(OH)_2$ suspension [1]. This rout is industrially used because of the availability of its raw materials, high yield as well as simplicity and low cost of production. Ultra-fine $CaCO_3$ could also be produced using the wet carbonation by adding surfactants and controlling the precipitation conditions, such as initial $Ca(OH)_2$ concentration, $CO_2$ flow rate, temperature, pH and stirring rate [2]. Anionic surfactants such as fatty acids and their sodium salts are widely used in industry for economic reasons. Such surfactants influence the $CaCO_3$ nucleation, crystal growth, grain shapes, and consequently, control the formation of crystal phases that are not usually stabilized under natural environments. They can bind with certain crystal planes during crystal growth, thereby modifying the particles morphology [3, 4]. The knowledge on the minimum quantity of surfactant that could produce the desired precipitate is of importance, because the excess of surfactant could deteriorate the physical and chemical properties of the hosting matrix [5]. The goal of the present study is to prepare ultra-hydrophobic $CaCO_3$ nanoparticles using adjusted wet carbonation rout. Sodium oleate was used during preparation to inhibit crystal growth and modify the hydrophobic property of the $CaCO_3$. The effect of sodium oleate on the particle size, morphology, surface potential, thermal decomposition and hydrophobicity were investigated.

## 2. Experimental

### 2.1. Materials

Analytical grade sodium oleate ($C_{18}H_{33}NaO_2$, 82+ % oleic acid, Sigma), calcium oxide (CaO, 97+ % on dry substance, Acros Organics), carbon dioxide gas ($CO_2$ gas, 99+ %, Air Liquide) and monodistilled water were used to prepare $CaCO_3$ particles.

### 2.2. Method

Unmodified $CaCO_3$ was prepared from CaO lime and $CO_2$ gas via the wet carbonation rout. The preparation conditions, initial CaO concentration, $CO_2$ flow rate and temperature were adjusted at 1 M CaO, 100 mL.min$^{-1}$ $CO_2$ flow rate and 25°C, respectively. For oleate-modified $CaCO_3$, 2 wt% sodium oleate based on the theoretical $CaCO_3$ weight was added to the monodistilled water. In practices, the preparation of $CaCO_3$ was carried out in a plastic flask. The required amount of CaO reagent was slaked in the monodistilled water containing sodium oleate then the obtained lime was cooled to 25°C. After cooling, the pure $CO_2$ gas was blown into the lime milk from the bottom of the plastic bottle under vigorous stirring. The pH value of the reaction solution was monitored online using a pH meter (Jenway 3305). When the pH value decreased to 9 the reaction was completed and then the $CO_2$ flow was stopped. The slurry was washed, filtered and dried at 120°C in an oven for 24 hrs to obtain the $CaCO_3$ powder.

### 2.3. Characterization

The crystalline phase of the prepared $CaCO_3$ was characterized using X-ray diffraction (XRD, Bruker axs D8, Germany) with Cu-K$\alpha$ ($\lambda$ = 1.5406 Å) radiation. The morphology information was investigated using a high resolution transmission electron microscope (TEM, Jeol, JEM-2010, Japan). Bonding structures were analyzed using Fourier transform infrared spectrometer (FTIR-460 plus, JASCO model 6100, Japan). Zeta surface potential of the $CaCO_3$ particles in suspension was measured using a Zeta meter (Malvern Instrument Zetasizer 2000). The thermal decomposition and weight loss was performed by thermogravimetric analysis (TA instrument, TGA Q5000, USA). The samples were dried isothermally at 55°C for 20 min before heating from 55 to 1000°C at a heating rate of 10 °C.min$^{-1}$ under Air atmosphere (50 mL.min$^{-1}$). The water contact angle (WCA) was measured with a Kruss DSA-100 contact angle analyzer. The measurements were performed on the prepared powders compressed into discs using 5 μL water droplet volume and the contacted angle was determined from the profile of the droplets that were fully separated from the pump syringe needle tip. The discs were prepared by compression under controlled conditions: 100 mg of the sample and a pressure of 107 Pa, in a typical IR die.

## 3. Results

X-ray diffraction patterns of the unmodified and oleate-modified $CaCO_3$ are shown in figure 1. XRD results indicate that the overall crystalline structure and phase purity of the $CaCO_3$ particles were obtained. XRD pattern exhibits the characteristic reflection of rhombohedral calcite (d-spacings/Å: 3.85, 3.04, 2.85, 2.49, 2.28; corresponding to hkl: 012, 104, 006, 110, 113, respectively) [6]. All the relative sharp peaks could be indexed as the typical calcite phase of $CaCO_3$ (JCPDS 88-1808). No characteristic peaks of other impurities were observed, which indicated that the products have high purity.

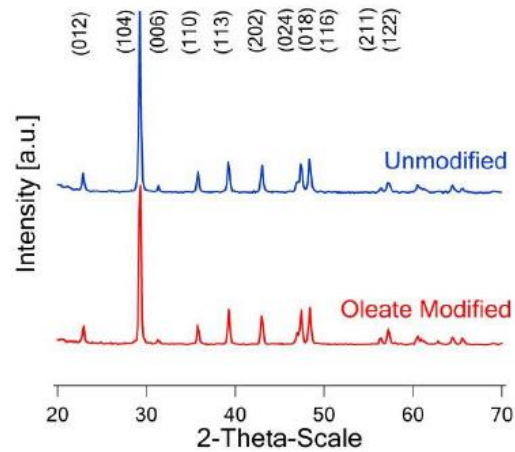

Figure 1. XRD patterns of the prepared calcium carbonates

FT-IR spectra of the unmodified and oleate-modified $CaCO_3$ are shown in figure 2. The characteristic absorption peaks of calcite $CaCO_3$ are stretching vibrations of the C—O approximately at 1420 cm$^{-1}$ and bending vibrations of the C—O approximately at 872 and 710 cm$^{-1}$. The combination of the three peaks at 1420, 872 and 710 cm$^{-1}$ appears at 2515 cm$^{-1}$. The broad absorption peaks at 3445 cm$^{-1}$ are assigned to stretching vibration and asymmetric stretching vibration of O—H bond and it can be attributed to the presence of absorbed water and hydroxyl groups on the surface of CaCO3 particles. The peaks around 2355 cm$^{-1}$ are attributed to carbon dioxide in the atmosphere. Comparing with the literature [6] the characteristic IR spectra indicate that all the prepared samples are typically calcite crystals. Comparing the unmodified and oleate-modified samples, a shoulder at 1615 cm$^{-1}$ and peaks at 2955, 2925 and 2885 cm$^{-1}$, are observed for oleate-modified $CaCO_3$. The shoulder at 1615 cm$^{-1}$ corresponds to the appearance of a carboxylic salt, indicating that oleate has been attached to the surface of $CaCO_3$ via an ionic bond. The peaks at 2955, 2925 and 2885 cm$^{-1}$, are ascribed to the long alkyl chain of oleate (H—C—H stretching region) and prove the presence of oleate at the surface of $CaCO_3$ [7]. The XRD and FT-IR results indicate that no characteristic absorption of other phases is observed meaning that sodium oleate has no significant effect on $CaCO_3$ polymorphism.

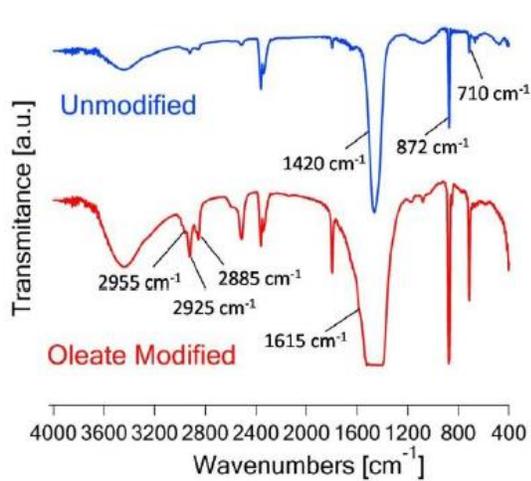

Figure 2. FT-IR spectra of the of the prepared calcium carbonates

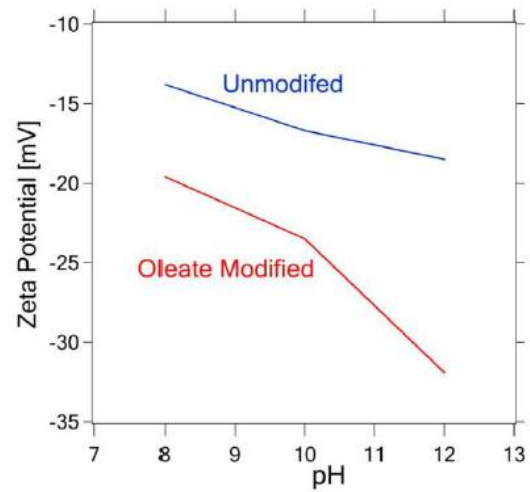

Figure 3. Zeta potentials of the prepared calcium carbonates

Zeta potentials of unmodified and oleate-modified $CaCO_3$ were determined are shown in figure 3. The Zeta potential measurements show that the particle surface charge decreases from -14.5 to -22 mV with addition of 2 wt% sodium oleate. The change in surface potential of $CaCO_3$ is due to adsorption of oleate on the surface of the precipitated particles.

Figure 4 shows TEM and water contact angle images of unmodified and oleate-modified $CaCO_3$. The results show that the use of 1 M CaO concentration and 100 mL.min$^{-1}$ forms 1.5-2.5 µm length scalenohedral particles. The average water contact angle (WCA) of these particles is about 25°. Upon addition of 2 wt% oleate, the average particle size of $CaCO_3$ reduced to 45 nm rhombohedral particles and the average water contact angle (WCA) increased to 130°.

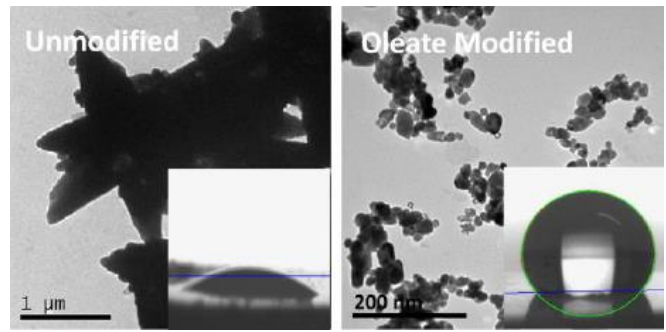

*Figure 4. TEM and WCA images of the prepared calcium carbonates*

Figure 5 shows the thermal decomposition curves of the unmodified and oleate-modified $CaCO_3$. The mass loss measurements revealed that the amount of adsorbed oleate reaches up to 75 % and the temperature at the maximum rate of decomposition of the $CaCO_3$ decreased from 757 to 715°C with addition of 2 wt% sodium oleate.

# 4. Discussion

Sodium oleate plays an important role on preparation and surface modification of the $CaCO_3$ particles. The data extracted from XRD and TEM reveal those oleates can successfully inhibit crystal growth and reducing the particle size of $CaCO_3$. The FT-IR, Zeta potential, TGA along with the WCA measurement confirm that sodium oleate can successfully adsorb on the $CaCO_3$ particles surface and alternates the particles surface properties from hydrophilic to ultra-hydrophobic. In absence of surfactant and the use of 1 M CaO, and low $CO_2$ flow rate of 100 mL.min–1 forms 2 µm length scalenohedral particles. Excess of $Ca^{2+}$ in the bulk solution and a low $CO_2$ flow rate elongate the reaction and growth time of precipitated particles and lead to microsize particle. Up on addition of 2 wt% sodium oleate, the particle size is reduced. This could be explained by that sodium oleates can be situated at the gas–liquid interface. Consequently they increase the stability of $CO_2$ bubbles and prevent their aggregation. The maintenance of $CO_2$ bubbles enhances the mass transfer of $CO_2$ into solution and increases the $CO_3^{2-}$:$Ca^{2+}$ ionic ratio and the nucleation rate. Moreover, the ability of oleate's carboxylic groups (–COO–) to bind effectively to the $Ca^{2+}$ ions on the crystals surface with ionic bonds can efficiently neutralize the positive charges on the crystals surface, decrease the surface charge to more negative values and inhibit the crystals growth [8]. The decrease of the particle size of the oleate-modified $CaCO_3$ causes the observed decrease of the thermal stability (temperature at maximum rate of decomposition) compared with unmodified $CaCO_3$ [9].

## 5. Conclusion

Ultra-Hydrophobic $CaCO_3$ particles were directly prepared via carbonation of CaO lime slurry in the presence of sodium oleate at room temperature. The sodium oleate solution was used in a precipitation process to control the particle size, morphology and to modify the surface characteristics of $CaCO_3$ particles simultaneously. The results from this work prove that coating of oleate on $CaCO_3$ particles was successfully done. Addition of oleate helps in reducing the particle size from 2 μm length scalenohedral particles to 45 nm rhombohedral particles and the water contact angle increases from 25° to 130° with addition of 2wt% oleate.